\documentclass[aps, pra,twocolumn,groupedaddress]{revtex4-2}
\usepackage{CJK}
\usepackage{color}
\usepackage{graphicx}
\usepackage{dcolumn}
\usepackage{subfigure}
\usepackage{graphics}
\usepackage{amsmath}
\usepackage{amsfonts}
\usepackage{amssymb}
\usepackage{epsf}
\usepackage{bm}

\begin{document}
\begin{CJK*}{GB}{ } 

\title{Coherent excitation of three-atom entangled states near a two-body F\"orster resonance}
\author{Tomohisa Yoda, Emily Hirsch, Jason Montgomery, Dilara Sen, and Aaron Reinhard}

\address{
Department of Physics, Kenyon College, 201 North College Rd., Gambier, Ohio 43022, USA \\
}

\date{\today}
\begin{abstract}
We experimentally study three-body energy exchange during Rydberg excitation near a two-body F\"orster resonance.  By varying the excitation pulse duration or Rabi frequency, we coherently control the excitation of three-atom entangled states.  We prove coherence using an optical rotary echo technique, and compare with a model for excitation in a three-atom basis.  Our results suggest a robust way to implement a three-body entangling operation.
\end{abstract}

\maketitle		
\end{CJK*}

\section{INTRODUCTION}

Interactions among ultracold atoms have enabled an explosion of progress in fundamental physics~\cite{Bloch} and quantum technologies~\cite{Saffman_Rev}.  Close-range interactions in quantum degenerate gases have revealed exotic phases of matter like Effimov states~\cite{Kraemer} and the Tonks-Girardeau gas~\cite{Kinoshita}.  They have also given insight into quantum dynamics, including phase transitions~\cite{Greiner,Hadzibabic}, wavepacket transport~\cite{Mandel}, Anderson localization~\cite{Kondov}, and quantum thermalization~\cite{Kinoshita2,Gring}.  Ultracold Rydberg atoms are particularly useful because of their long-range couplings~\cite{Saffman_Rev}.  Dipole-dipole interactions in Rydberg systems have been used to create neutral atom quantum gates~\cite{Isenhower,Bluvstein,Ma}, flexible quantum simulators~\cite{Ebadi,Scholl}, single photon sources~\cite{Dudin2,Ripka}, and single photon switches and transistors~\cite{Baur,Tiarks}.  These long-range interactions have also opened new avenues to study few-body physics, including Rydberg molecules~\cite{Bendkowsky,Booth}, facilitated excitation~\cite{Ates,Amthor}, and few-photon optical nonlinearities~\cite{Pritchard,Liang,Liang2}.

There has been significant recent interest in non-radiative, dipolar energy transfer.  This transfer occurs most readily at F\"orster resonance, or a near-degeneracy between multi-atom Rydberg states.  The process is reminiscent of fluorescence resonance energy transfer (FRET), proposed by F\"orster to explain energy transport in biological systems~\cite{Forster,Jang,Sener}.  State-changing energy transfer has been studied for several decades near two-atom F\"orster resonances, but beyond two-body effects have been difficult to confirm \cite{Anderson,Mourachko,Sun,Richards}.  This work was recently extended to three- and four-atom F\"orster resonances~\cite{Gurian,Faoro,Tretyakov,Beterov,Ryabtsev,Liu}, and the Borromean nature of one such resonance was demonstrated~\cite{Faoro,Tretyakov}.  Energy exchange near few body resonances may shed light on many-body localization~\cite{Nandkishore} and quantum thermalization~\cite{Liu}. It has also been proposed as an entangling operator for quantum computation and simulation~\cite{Tretyakov,Beterov,Ryabtsev}.  However, coherent dipole-dipole energy exchange has not been previously observed at the three-atom level.

In most recent work, the energies of Rydberg states were manipulated via the DC Stark effect to fulfill a precise resonance condition.  Here, we demonstrate controllable three-body energy exchange near a two-body F\"orster resonance.  This process, $2 \times nD_{5/2} \rightarrow (n-2)F_{7/2} + (n+2)P_{3/2}$, is nearly resonant in rubidium near $n=43$.  We see strong evidence of coherent excitation of three-atom entangled states in zero applied field.  The mechanism we study is particularly robust, since it is insensitive to the precise value of the energy defect.

We study the process proposed by Pohl and Berman in Ref.~\cite{Pohl}.  Consider Rydberg excitation of three atoms from a ground state, $\vert g \rangle$, to a target state, $\vert d \rangle$, via a resonant optical pulse.  Near F\"orster resonance, two- particle states $\vert dd \rangle$ are coupled to nearby product states, $\vert pf \rangle$ and $\vert fp \rangle$, and the energy defect $\Delta E =2 \times E_{d} - (E_{p}+E_{f}) \approx 0$.  An optical pulse can drive transitions between the three-atom ground state $\vert ggg \rangle$ and a triply-excited state, $\vert M_3 \rangle$, through virtual levels whose populations follow the pulse envelope.  This entangled state has zero energy shift, and has the form $\vert M_3 \rangle = c_1 \vert dpf \rangle + c_2 \vert dfp \rangle + c_3 \vert pdf \rangle + c_4 \vert fdp \rangle + c_5 \vert pfd \rangle + c_6 \vert fpd \rangle$, where $c_i$ are probability amplitudes.  Given appropriate values for the pair-state detuning, pulse duration, and Rabi frequency, $\vert M_3 \rangle$ should be excited with high probability~\cite{Pohl}.

\begin{figure}[htp]
\centerline{ \scalebox{0.28} {\includegraphics{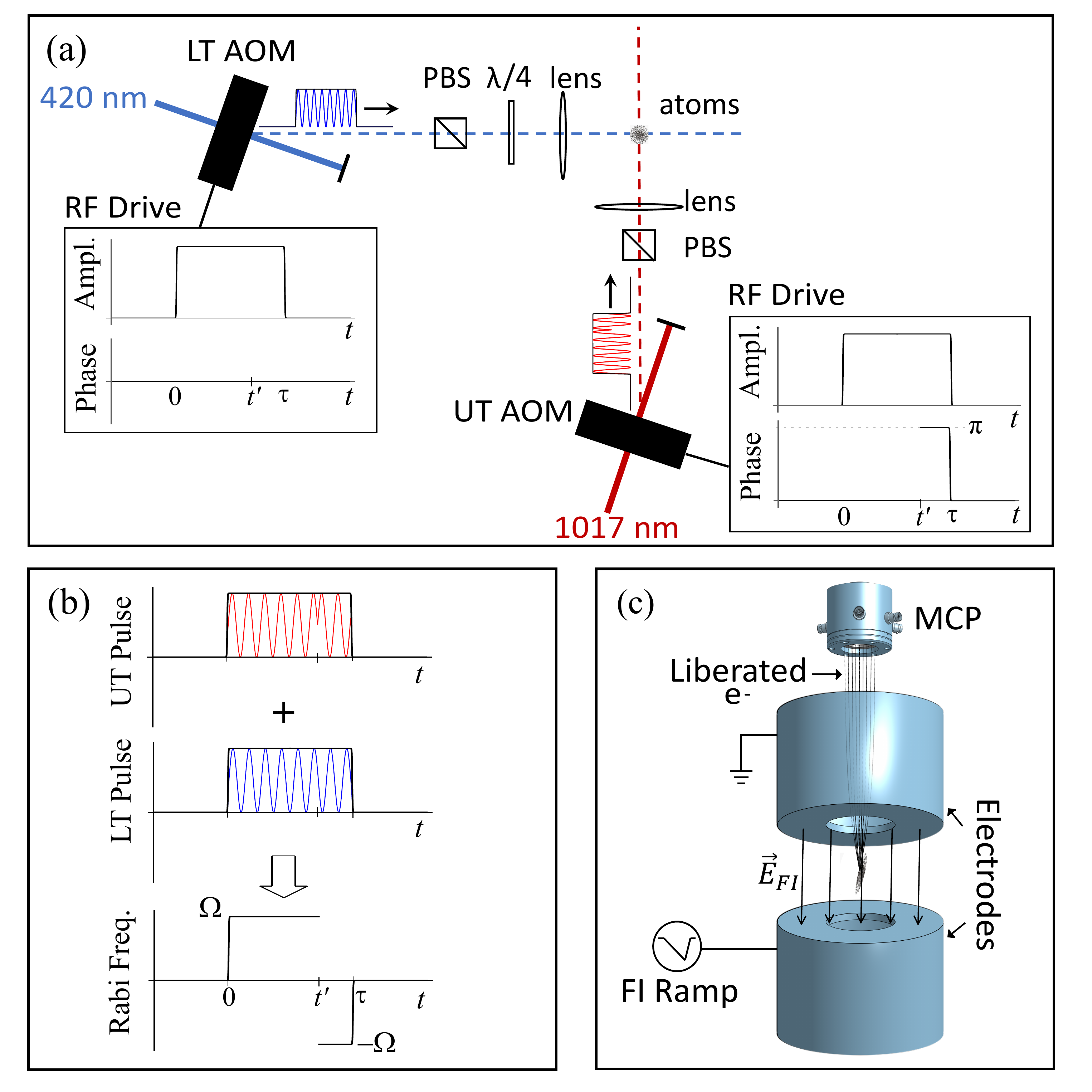}}} \caption{ \label{Fig1}  (a) Schematic of the experiment.  Optical pulses are created by amplitude modulating radiofrequency (RF) signals driving acousto-optic modulators (AOMs).  The phase of the RF signal driving the lower transition (LT) AOM is constant. At time $t'$, the phase of the RF signal driving the upper transition (UT) AOM is shifted by $\pi$.  This (b) flips the phase of the UT optical pulse, and the effective Rabi frequency  switches from $\Omega$ to $-\Omega$, where $\Omega = \Omega_L \Omega_U / 2 \Delta$. (c) Side view of the apparatus, illustrating state-selective field ionization. }
\end{figure}

Previous work has shown that when one excites $^{85}$Rb atoms to $nD_{5/2}$ Rydberg states ($\vert d \rangle$) near $n=43$, a large fraction of Rydberg atoms are detected in $(n+2)P$ ($\vert p \rangle$) and $(n-2)F$ ($\vert f \rangle$) states immediately after excitation~\cite{Reinhard,Younge,Kondo,Eder}.  We recently developed a method to determine if the energy exchange is two- or three-body in nature~\cite{Eder}.  In the present work, we use this technique to establish control of excitation into the triply excited states, $\vert M_3 \rangle$, as we vary excitation pulse duration or Rabi frequency.  We employ an optical rotary echo~\cite{Solomon,Raitzsch,Younge2,Thaicharoen,Hernandez} to prove coherence, and we find good agreement between our data and the model of Ref.~\cite{Pohl}.  The coherent signal remains very strong, even when exciting multiple excitation domains with a spatially inhomogeneous laser, in a disordered density distribution, and with uncontrolled $m_j$.  Therefore, excitation of $\vert M_3 \rangle$ might be useful as a three-body entangling operator in situations with less-than ideal control over experimental parameters.  This is promising, because systems with single-atom control are difficult to achieve.  Examples of technologies requiring three interacting atoms include Toffoli and Fredkin gates~\cite{Galindo,Shi}, or quantum simulations of exotic spin Hamiltonians~\cite{Peng,You,Luo}.  A Toffoli gate based on non-radiative three-body energy transfer, rather than the usual dipole blockade, was recently proposed~\cite{Beterov}.

\section{EXPERIMENTAL PROCEDURE}

Our setup is described elsewhere~\cite{Eder}.  Briefly, we collect ultracold $^{85}$Rb atoms in an optical dipole trap.  We control the ground state atom density by turning off the dipole trap beam and allowing the atoms to freely expand before they are excited to Rydberg states~\cite{Younge}.  We use a relatively low density:  $\rho \sim 5 \times 10^{10}$~cm$^{-3}$.  We apply coincident pulses of duration $\tau$ to drive the $5S_{1/2} \rightarrow 6P_{3/2} \rightarrow 42D_{5/2}$ transitions, with an intermediate state detuning $\Delta=50$~MHz.  Since $\tau \leq 2~\mu$s, the atoms are effectively frozen during Rydberg excitation.  The $\sigma +$ polarized lower transition beam is derived from a MOGLabs external cavity diode laser.  The $\pi$ polarized upper transition beam is derived from a MOGLabs cateye laser that is frequency stabilized to a pressure-tunable Fabry P\'erot cavity~\cite{Orr}.  The beams are perpendicular to each other and to the long axis of our dipole trap.  Pulses are created by amplitude modulating the radiofrequency (RF) signal driving acousto optic modulators (AOMs) in each beam, as shown in Fig.~\ref{Fig1}a.

We detect atoms using state-selective field ionization (SSFI) spectroscopy, outlined in Fig.~\ref{Fig1}c.  A high voltage ramp is applied to electrodes above and below the atom cloud, 50~ns after each excitation sequence.  Atoms with different binding energies will ionize at different electric fields, and the liberated electrons are detected by a dual stage microchannel plate detector (MCP).  For each of 1001 shots of our experiment, we use a pulse counter to record the number of excitations in each of two independent timing gates.  The ``P Gate'' counts atoms in $44P_{3/2}$, or $\vert p \rangle$, while the ``T Gate'' counts all Rydberg atoms.  From this data, we construct a ``sorted graph,'' or a 2D histogram of the total number of excitations as a function of the number in $\vert p \rangle$.  We fit each sorted graph to a linear function.  The slope tells us how many additional Rydberg excitations are created each time an atom is detected in $\vert p \rangle$.  A small value of slope indicates that the energy exchange is dominated a two-body process, since one additional Rydberg atom (in $\vert f \rangle$) is created for each atom in $\vert p \rangle$.  A large value of the slope indicates the presence of three-atom entangled states of the type $\vert M_3 \rangle$, because two additional Rydberg atoms (in $\vert d \rangle$ and $\vert f \rangle$) are created for each atom in $\vert p \rangle$~\cite{Eder}.

\begin{figure}[htp]
\centerline{ \scalebox{0.45} {\includegraphics{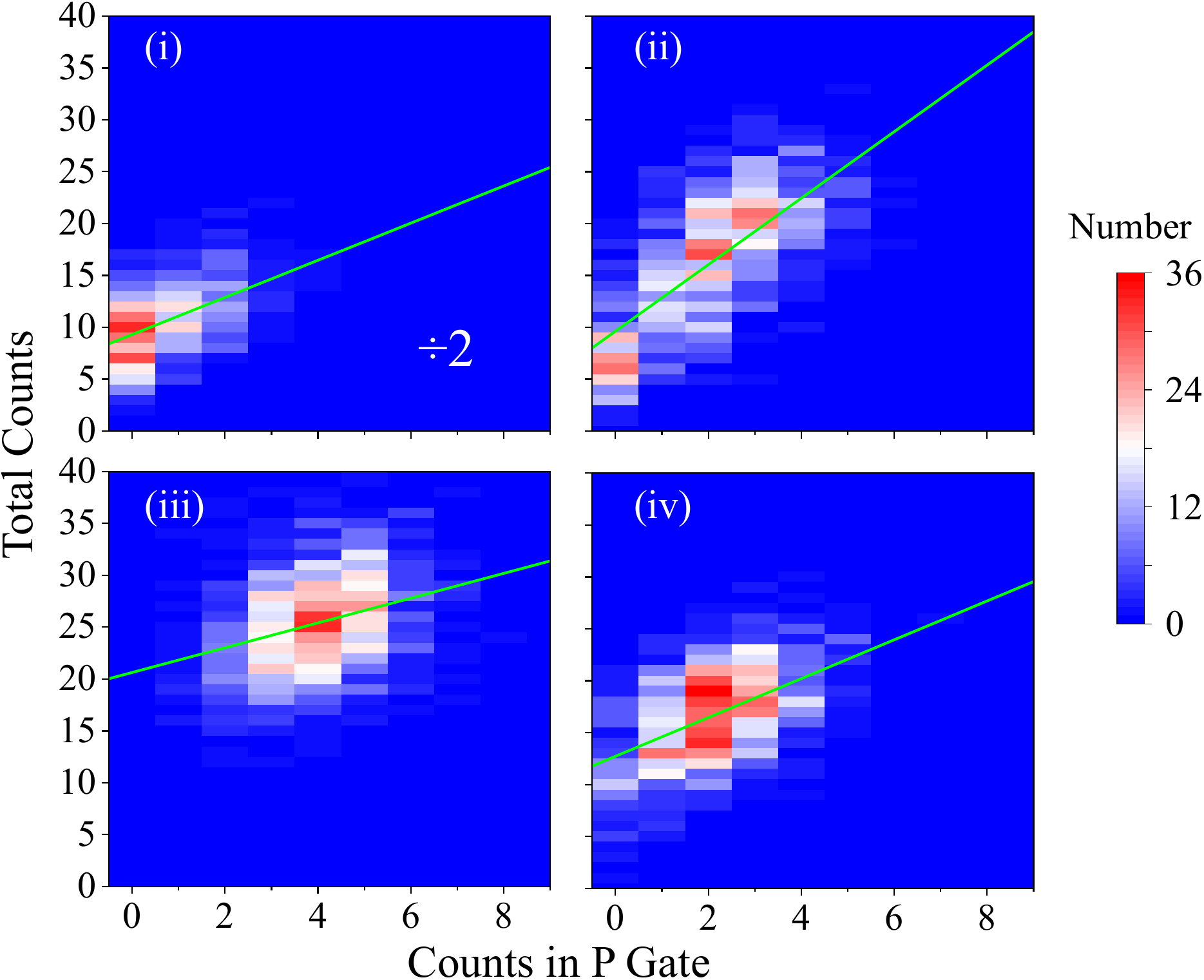}}} \caption{ \label{Fig2} Sorted graphs for excitation to the $42D_{5/2}$ state with fixed $\Omega = 1$~MHz with (i) $\tau=200$~ns, (ii) $\tau=500$~ns, (iii) $\tau=2000$~ns, and (iv) $\tau=500$~ns; phase flip at $t'=250$~ns (see Fig.~\ref{Fig3}).  These graphs show the total number of excitations, $N_T$, as a function of the number in $\vert p \rangle$, $N_P$.  The false color indicates how many of each $\{N_P, N_T\}$ were detected.  The green line is a fit, from which we extract the slope.  Since there is less spread in $\{N_P, N_T\}$ in panel (i), all values have been divided by 2 to match the common false color scale.}
\end{figure}

Example sorted graphs are shown in Fig.~\ref{Fig2} for fixed Rabi frequency $\Omega = \Omega_L \Omega_U / 2 \Delta = 1$~MHz.  The green lines are the least-squares linear fit to the data.  In panels (i) through (iii) the pulse duration, $\tau$, is increased from 200~ns to 500~ns to 2000~ns.  The slope of the sorted graphs clearly increase and then decrease.  This suggests that the mechanism causing energy exchange near F\"orster resonance is highly sensitive to pulse duration.

\section{RESULTS AND DISCUSSION}

\begin{figure}[htp]
\centerline{ \scalebox{0.45} {\includegraphics{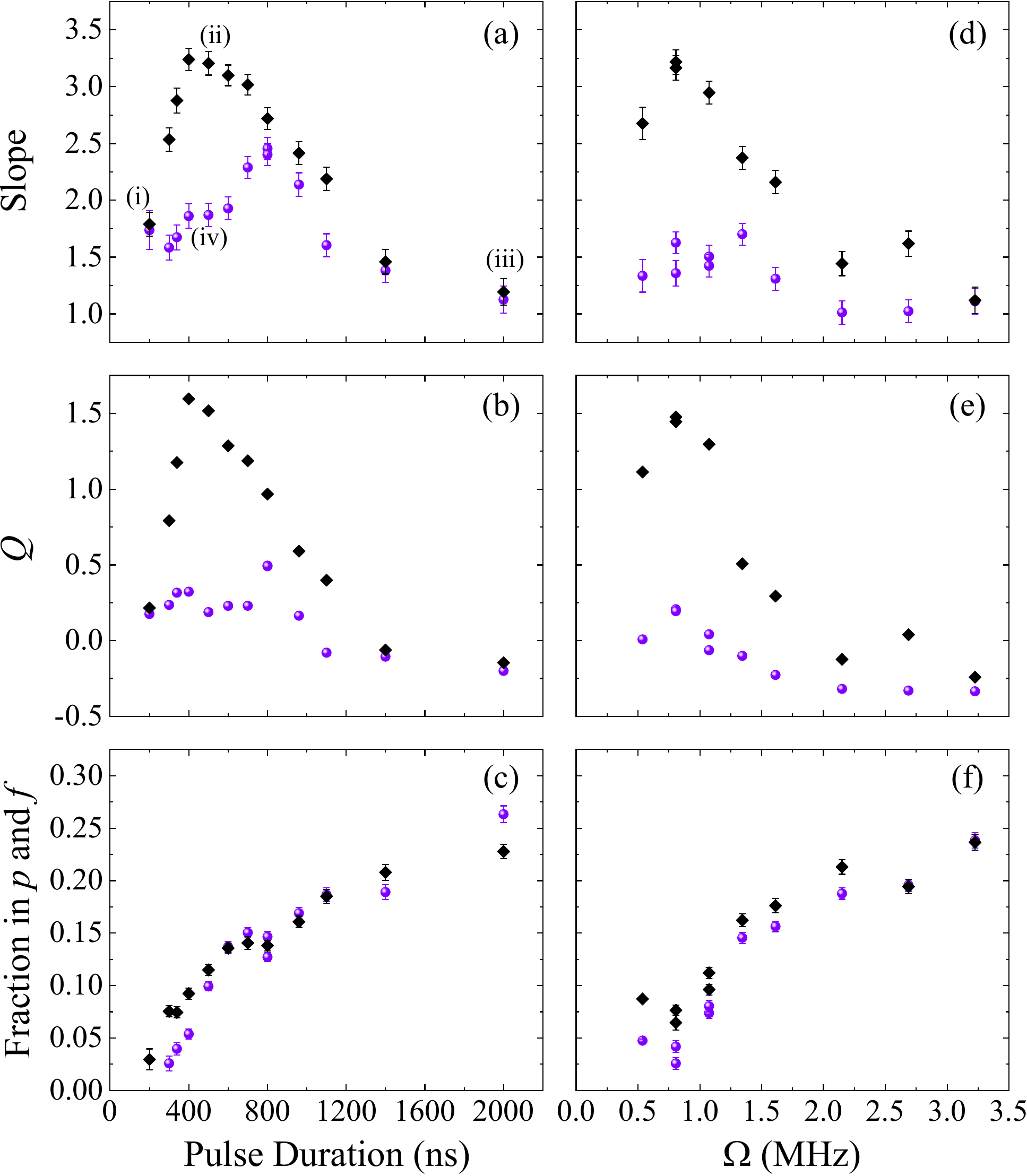}}} \caption{ \label{Fig3} Slopes of the sorted graphs, Mandel $Q$, and mixing fraction for atoms excited to the $42D_{5/2}$ state.  In panels (a)-(c) $\Omega= 1$~MHz, and $\tau$ is varied.  In panels (d)-(f) $\tau=400$~ns, and $\Omega$ is varied.  The black diamonds are for excitation with no upper transition phase flip.  The purple circles are for excitation with a phase flip at $t' = \tau / 2$.  Error bars are the one-sigma uncertainty resulting from a least-squares fit to our raw data.}
\end{figure}

To explore further, we measured the slopes of the sorted graphs as a function of $\tau$ for fixed $\Omega$ and as a function of $\Omega$ for fixed $\tau$.  The results are shown as black diamonds in the top two panels of Fig.~\ref{Fig3}.  We also plot two other quantities: the Mandel $Q$ parameter and the mixing fraction.  The Mandel $Q$ parameter represents the width of the distribution of number of excitations.  It is defined as $Q=\sigma^2 / \bar{N}_T-1$, where $\sigma^2$ is the dispersion and $\bar{N}_T$ is the mean number of total excitations.  The mixing fraction is defined as the fraction of Rydberg excitations found in $\vert p \rangle$ and $\vert f \rangle$~\cite{note}. It indicates the fraction of Rydberg excitation events that result in energy exchange, and increases monotonically with pulse duration and Rabi frequency.  In contrast, the slopes of the sorted graphs and the Mandel $Q$ parameter go through clear relative maxima.

To test the coherence of the process causing these maxima, we implemented an optical rotary echo technique~\cite{Solomon,Raitzsch,Younge,Thaicharoen}.  The sequence, shown schematically in Figs.~\ref{Fig1}a and b, is similar to rotary spin echo in nuclear magnetic resonance~\cite{Solomon}.  At a variable time, $t'$, within our upper transition excitation pulse, we reverse the phase of the RF signal driving an AOM.  This reverses the sign of the excitation Rabi frequency, $\Omega$.  Independent of the value of $\Omega$, the system should undergo reverse evolution and arrive back to its ground state at a time $2t'$, unless some dephasing has occurred.  The plots of slope, Mandel $Q$, and mixing fraction using a phase flip at half the pulse duration ($t'=\tau/2$) are shown in Fig.~\ref{Fig3} as purple circles.  A sorted graph with phase flip at $\tau/2$ is shown in panel (iv) of Fig. 2.  In both Figs.~\ref{Fig2}~and~\ref{Fig3} it is clear that, while the phase flip does not significantly change the fraction of atoms excited into product states, it dramatically reverses the evolution into multiparticle states that lead to large slope and $Q$.  Thus, we conclude that the excitation of these multiply excited states is coherent over most of the range of $x$-axis values.

\begin{figure}[htp]
\centerline{ \scalebox{0.45} {\includegraphics{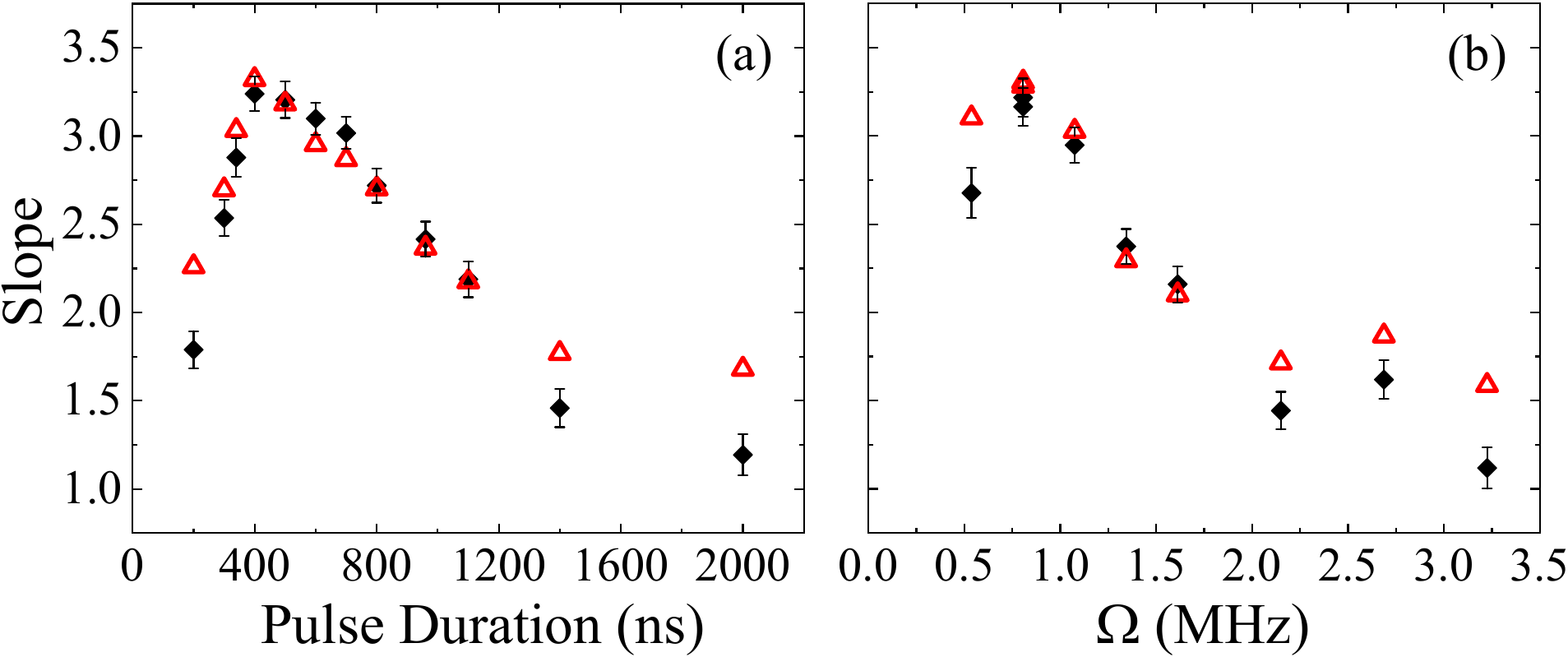}}} \caption{ \label{Fig4} Slopes of the sorted graphs with no upper transition phase flip from Fig.~\ref{Fig3} (black diamonds) along with the results of a Monte Carlo model (red triangles).  The model accounts for the effect of atom number fluctuations and non-unity detector efficiency on the slopes.  In panel (a) $\Omega = 1$~MHz, and $\tau$ is varied and in panel (b) $\tau=400$~ns, and $\Omega$ is varied.}
\end{figure}

To determine the nature of the multiply-excited states, we compare the measured values of the slopes of the sorted graphs with the results of a Monte Carlo simulation.  We account for the effects of non-unity detector efficiency and shot-to-shot fluctuations in excitation number on the slopes. We assume each excitation event results in either one atom in $\vert d \rangle$ or three atoms, one each in $\vert p \rangle$, $\vert d \rangle$, and $\vert f \rangle$.  The model then predicts a value for the slope, given the $Q$-value and mixing fraction present in the experiment  (see Ref.~\cite{Eder} or the appendix for details).  Figure~\ref{Fig4} shows the no-phase flip data from Figs.~\ref{Fig3}a and d, along with the predictions of our Monte Carlo model. The peaks in our data are consistent with the creation of triply-excited states.  In the case of a phase flip at $\tau/2$, there is no significant peak in the slope, as seen in Fig.~\ref{Fig3}.  In this case, triply-excited states are not created with high probability.

To determine if the peaks in Figs.~\ref{Fig3} are due to excitation of $\vert M_3 \rangle$, we implemented the model described in Ref.~\cite{Pohl}.  This model describes excitation in a three atom basis, and we modified it by adding always resonant ``hopping'' couplings $\vert d \rangle \leftrightarrow \vert f \rangle$ and $\vert d \rangle \leftrightarrow \vert p \rangle$.  We numerically solved the time-dependent Schr\"odinger equation (TDSE) for various coupling strengths $V_{i,j}$ (or distances $r_{i,j}$), where $\{i,j\} \in \{1,2,3\}$.  We first placed the atoms on an equilateral triangle and calculated the maximum probability to find the system in a state with one atom each in $\vert d \rangle$, $\vert p \rangle$ and $\vert f \rangle$ as $\tau$ was varied.  The probability to create triply-excited states is a sharply peaked function of the triangle's side length, $r$, with a maximum value at $r_{\rm{max}}=3.25~\mu$m and full width at half maximum (FWHM) $1.07~\mu$m for $\Omega=1$~MHz.  We then solved the TDSE, averaging over random atom placements within the experimental distribution of Rabi frequencies.  We placed the atoms inside a shell of radius $r_{\rm{max}}$ and width equal to the FWHM.  This is the volume inside of which the excitation of triply-excited states will be most probable.  We recorded the probability to detect one atom each in $\vert d \rangle$, $\vert p \rangle$, and $\vert f \rangle$. For $\Omega=1$~MHz there are, on average, about 8 atoms inside of the shell.  We accounted for the fact that there are $N=\binom{8}{3}$ possible triples which could be excited within this volume.  Since each of these excitation channels is independent and do not interfere, we multiplied the triple excitation probability by the appropriate value of $N$ for each Rabi frequency.

\begin{figure}[htp]
\centerline{ \scalebox{0.4} {\includegraphics{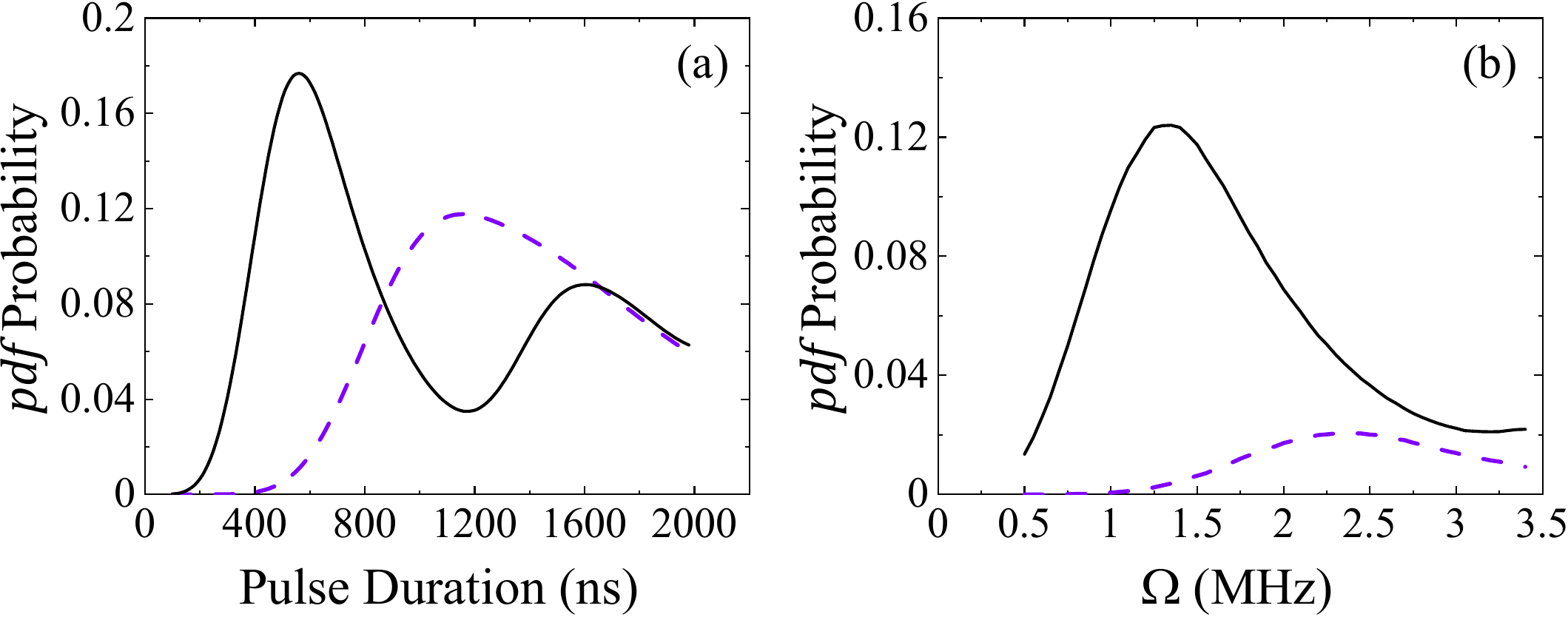}}} \caption{ \label{Fig5} Calculated probability to excite a state with one atom each in $\vert d \rangle$, $\vert p \rangle$, and $\vert f \rangle$, using to the model in Ref.~\cite{Pohl}.  The black solid line is for no phase flip and the purple dashed line is for a phase flip at $t'=\tau/2$.  Panel (a) is for $\Omega=1$~MHz and panel (b) is for a pulse duration of 400~ns.}
\end{figure}

The results of the calculation are shown in Fig.~\ref{Fig5}.  The black curves are for no Rabi frequency inversion and the purple dashed curves are for an inversion at $t'=\tau/2$.  It is notable that the Rabi oscillation shows only moderate damping, despite the fact that we average over many random placements in an inhomogeneous intensity distribution. For no rotary echo, the maximum probability occurs at approximately the same pulse duration (panel a) as the data in Fig.~\ref{Fig3}a, but at a larger Rabi frequency (panel b) than the data in Fig.~\ref{Fig3}d.  For Rabi frequency inversion at $t'=\tau/2$ the calculated maxima occur at even longer times (panel a) and higher Rabi frequencies (panel b).  In Figs.~\ref{Fig3}a and d we don't see strong evidence of these peaks.  Any disagreements between theory and experiment have a common cause. Whenever we drive our atoms strongly (long pulses or large Rabi frequency), excitation of doubly-excited states containing terms like $\vert pf \rangle$ or $\vert fp \rangle$ dominates over the coherent excitation of $\vert M_3 \rangle$.  We increase the probability of exciting close pairs into unshifted branches of the molecular potential curves~\cite{Kondo,Derevianko,Eder}.  Whenever this happens, it causes the slope and Mandel $Q$ to decrease.  Nonetheless, the agreement between the data in Fig.~\ref{Fig3} and the shapes of the calculated curves in Fig.~\ref{Fig5} indicates that we are exciting the entangled state $\vert M_3 \rangle$.

Interestingly, if we repeat the experiments described in this paper when exciting to $43D_{5/2}$ states, we don't see any coherent features in our graphs of slope vs. pulse duration or Rabi frequency.  Since the magnitude of the F\"orster defect is much smaller for $n=43$ than for $n=42$ (-11 vs. -97~MHz), one would naively expect a higher probability to excite $\vert M_3 \rangle$.  However, this is not the case.  The reason is that, for $n=43$, molecular potential branches cross zero at larger separations, and with larger overlap with the asymptotic state~\cite{Eder}.  Therefore, atoms are readily excited into two-body states featuring terms like $\vert pf \rangle$ or $\vert fp \rangle$.  The unfavorable molecular potential curves dominate the dynamics, no matter our choice of experimental parameters.

\begin{figure}[htp]
\centerline{ \scalebox{0.415} {\includegraphics{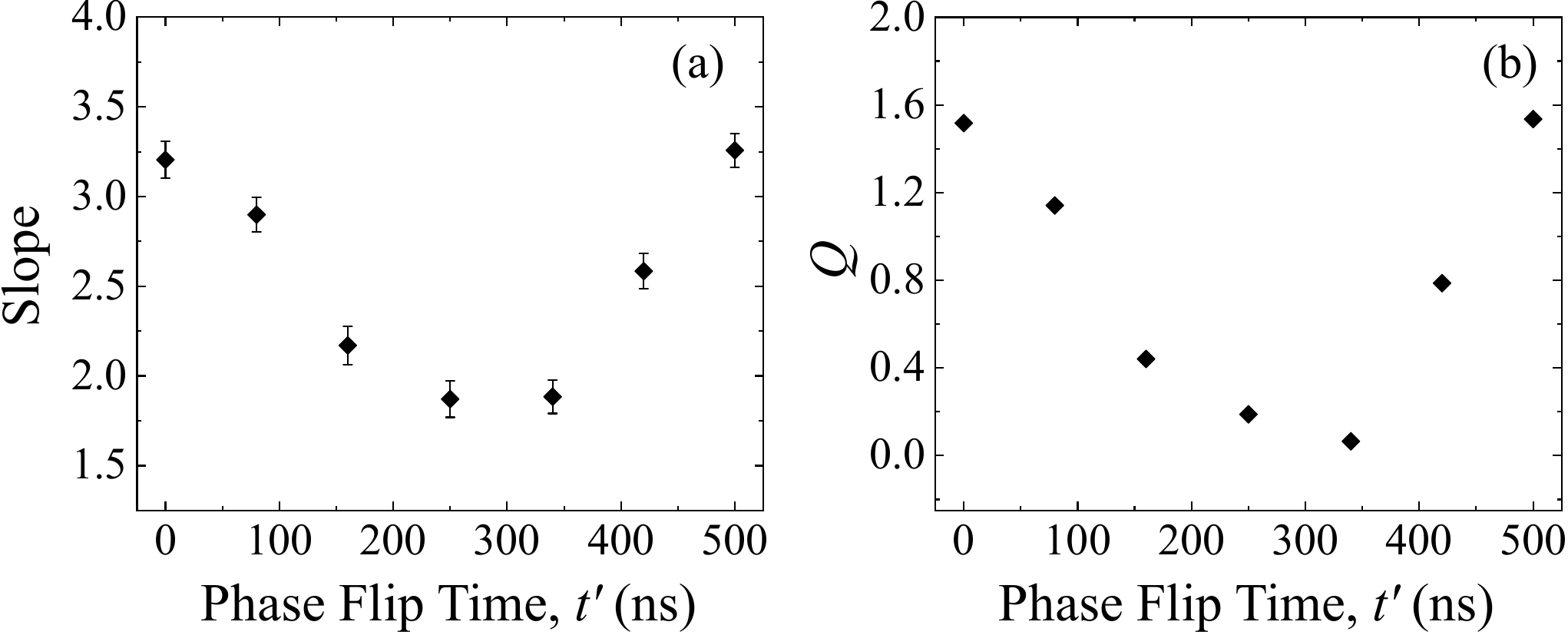}}} \caption{ \label{Fig6} Slopes of the sorted graphs (a) and Mandel Q (b) with fixed $\Omega = 1$~MHz and $\tau = 500$~ns.  The time, $t'$, of the upper transition phase inversion is varied.}
\end{figure}

To further demonstrate the coherence of the evolution into $\vert M_3 \rangle$, we vary the echo time, $t'$, for fixed $\Omega = 1$~MHz and $\tau = 500$~ns.  Thus, we should excite $\vert M_3 \rangle$ for a time $t'$ and reverse the evolution for a time $\tau - t'$.  The slopes of the sorted graphs and the Mandel $Q$ are shown in Fig.~\ref{Fig6}.  The clear relative minimum gives strong evidence of the reversibility of the evolution.  Note that in this graph, as in the previous graphs, the Mandel $Q$ closely follows the behavior of the slopes of the sorted graphs.  Excitation events may always produce single $\vert d \rangle$ excitations or pairs of atoms in $\vert p \rangle$ and $\vert f \rangle$ (due to short-range molecular potential zero crossings).  However, as the probability to excite three-body entangled states increases, the fluctuations in excitation number also increase.  This broadens the excitation number distributions.

\section{CONCLUSION}

We have demonstrated coherent excitation of three-atom entangled states near F\"orster resonance.  This is the first observation of coherent dipole-dipole energy exchange beyond two atoms.  Coherence was proven using an optical rotary echo technique.  We used a Monte Carlo method to demonstrate that coherent features in our data are consistent with the excitation of triply excited states.  Finally, we showed that our data is well-described by the model of Ref.~\cite{Pohl}.  The demonstrated coherence of $\vert M_3 \rangle$ in a bulk gas suggests that this state might find application in quantum technologies, in situations where experimental conditions cannot be carefully controlled.

{\em Acknowledgements.}  The authors acknowledge valuable input from Paul Berman, Georg Raithel, Smitha Vishveshwara, and David Weiss. This work was supported by NSF Grants PHY-1745628 and PHY-2204899.

\appendix*
\section{DETAILS OF THE MONTE CARLO SIMULATION}
The slopes of the sorted graphs provide information about the mechanism behind dipole-dipole energy transfer into $\vert p \rangle$ and $\vert f \rangle$.  A lower value of the slope indicates that a two-body mechanism causes the transfer (one additional Rydberg atom for each atom in $\vert p \rangle$), while a larger value indicates that a three-body mechanism is responsible (two additional Rydberg atoms for each atom in $\vert p \rangle$).  However, two factors make it difficult to directly interpret the slopes: non-unity detector efficiency and shot-to-shot fluctuations in the number of excitation events.  These effects can combine to change number of Rydberg excitations \textit{detected} in the two counting gates, for a given \textit{true} number of excitations created.

To model the impact of these factors on our slope data, we use a Monte Carlo simulation.  Each run of the progam draws a random number of excitation events from a Gaussian distribution, whose width is chosen to reproduce the experimentally measured value of the Mandel $Q$ parameter (or level of excitation fluctuation).  We randomly assign each excitation event to either a single atom in the (target) $\vert d \rangle$ state, or three atoms, one each in $\vert d \rangle$, $\vert p \rangle$, and $\vert f \rangle$.  The probability to excite a three-atom state is given by the experimentally measured mixing fraction.  After every excitation event, the program decides if each Rydberg count is recorded, according to our microchannel plate's detector efficiency.

For each Monte Carlo point in Fig.~4, we run the simulation $5 \times 10^7$ times, and plot the total number of Rydberg excitations as a function of the number in $\vert p \rangle$.  We fit the simulated sorted graph to a line and extract the slope, just as in the experiment.  Therefore, the Monte Carlo model predicts a value of the slope, given a three-atom model for state mixing and the experimentally measured mixing fraction and Mandel $Q$.  As shown in Fig.~4, the coherent features in our experimental data are consistent with a three-atom model for excitation into product states.


\begin{thebibliography}{99}
\bibitem{note} The mixing fraction is calculated as twice the integral of the $p$-peak in the SSFI spectrum, divided by the total integral.
\end{thebibliography}
\end{document}